\documentstyle[12pt]{article}
 \begin{document}
\medskip
\begin{center} \large
               {\bf Irving Kaplansky and Supersymmetry\footnote{Based on remarks at the Irving Kaplansky Memorial at MSRI, Berkeley, Ca, February 23, 2007 }\\
	\normalsize
		   
			    \bigskip \bigskip                   Peter G.O. Freund\footnote{freund@theory.uchicago.edu}\\
 \em{Enrico Fermi Institute and Department of Physics\\
 University of Chicago, Chicago, IL 60637}}\\

\bigskip \bigskip

 \bf{Abstract}\\
 \end{center}

Remarks at the Irving Kaplansky Memorial about a collaboration during the early period of the renewal of contacts between mathematicians and theoretical physicists.
 
\bigskip \bigskip\bigskip \bigskip

I arrived in Chicago some two decades after Irving Kaplansky, and I met Kap, as we all called him, shortly after my arrival here. We became friends later, in 1975, while collaborating on a paper on supersymmetry.

Lie superalgebras, graded counterparts of ordinary Lie algebras, play a central role in string theory and other unified theories. A classification of the simple ones was of essence. I took some initial steps, but the real work started when Yitz Herstein put me in touch with Kap. At first, communication was not easy. We couldn't quite make out each other's reasoning, much as we agreed on results. It didn't take long however, to get used to the other's way of looking at things. Mathematicians and physicists think in similar ways after all, all that was needed was a dictionary.

This was during the early phase of the rapprochement between mathematics and theoretical physics. After the glorious first half of the twentieth century --- when the likes of Poincar\'{e}, Hilbert, Weyl, von Neumann, \'{E}lie Cartan, Emmy Noether, and others made major contributions to the then new physical theories of general relativity and quantum mechanics, while physicists like Jordan, Dirac, Casimir and Feynman made major contributions to mathematics --- physics entered a period best described as phenomenological. During this period, some advanced complex function theory aside, very little modern mathematics was drawn on. To give you an idea, when in his celebrated "Eightfold Way" paper,  Murray Gell-Mann wrote down a basis of the three-dimensional representation of the ${\em su(3)}$ Lie algebra, this was heralded by physicists as a great mathematical feat. "Imagine, he found a $3\times 3$ generalization of the famous $2\times 2$ Pauli matrices," is what most people said. To get there, Murray had consulted with Block and Serre! 

It was in the fields of supersymmetry and gauge theory that the initial steps in modern mathematical physics were taken. This convergence of the paths of mathematics and of theoretical physics is typical of times when major new physical theories --- gauge theory and string theory in this case --- are being born.  The earliest example of such a convergence is the creation of calculus at the birth of Newton's mechanics and of his theory of gravitation. Weyl's spectacular work on group theory under the impact of the newborn quantum mechanics is another such example.

A few words about our joint paper \cite{FK} are in order here. In it we found all the infinite families of simple Lie superalgebras, as well as 17-, 31- and 40-dimensional exceptional ones. We also discussed real forms, and explained why supersymmetry can act on 4-dimensional anti-de Sitter but {\em not} on de Sitter space, a result essential for understanding why the remarkable duality discovered by Maldacena \cite{M} in the nineties, is of the AdS/CFT and not of the dS/CFT type. We were convinced that we had found all simple Lie superalgebras (as we actually had), but we lacked a proof of this fact. The proof came from the powerful independent work of Victor Kac \cite{K1}. Amusingly, in his beautiful proof, Kac somehow overlooked one of the exceptional superalgebras, namely the 31-dimensional superalgebra $G(3)$, whose Bose (even) sector consists of the ordinary Lie algebra $\em g_2 + sl(2)$, the only simple Lie superalgebra to have an exceptional ordinary Lie algebra as one of the two constituents of its Bose sector. I said "amusingly" above because, as I learned from Kap, in the classification of {\em ordinary} simple Lie algebras, in his extremely important early work, Killing had found  {\em almost all} of them, but he "somehow overlooked one," namely the exceptional 52-dimensional simple Lie algebra $F_4$, which remained to be discovered later by \'{E}lie Cartan. Apparently, $G(3)$ is the exceptional Lie superalgebra which carries on that curse of the ordinary exceptional Lie algebra $F_4$.
 
I mentioned the almost total lack of contact between theoretical physicists and mathematicians, when this work got going. It went so deep that in 1975 most physicists, if asked to name a great modern mathematician, would come up with Hermann Weyl, or John von Neumann, both long dead. Mathematicians had it a bit easier, for if they read the newspapers, they could at least keep track of the Nobel Prizes, whereas newspaper editors rarely treated Fields medal awards as "news fit to print." 

I recall that while standing by the state-of-the-art Xerox machine to produce some ten copies of our paper in about ... half an hour's time, I asked Kap, "Who would you say, is the greatest mathematician alive?" He immediately took me to task: my question was ill-defined, did I mean algebraist, or topologist, or number-theorist, or geometer, or differential geometer, or algebraic geometer, etc...  I replied that I did not ask for a rigorous answer, but just a "gut-feeling" kind of answer. "Oh, in that case the answer is simple: Andr\'{e} Weil," he replied, without the slightest hesitation, a reply that should not surprise anyone, who has heard today's talks. "You see," Kap went on, "We all taught courses on Lie algebras or Jordan algebras, or whatever. By contrast, Weil called all the courses he ever taught simply 'mathematics' and he {\em lived up} to this title." 

Kap went on to tell me about Weil's legendary first colloquium talk in Chicago. This was the first time I heard that very funny story. Weil had been recruited for the Chicago Mathematics Department by its chairman, Marshall Stone. With Stone sitting in the first row, Weil began his first Chicago colloquium talk with the observation, "There are three types of department chairmen. A bad chairman will only recruit faculty worse than himself, thus leading to the gradual degeneration of his department. A better chairman will settle for faculty roughly of the same caliber as himself, leading to a preservation of the quality of the department. Finally, a good chairman will only hire people better than himself, leading to a constant improvement of his department. I am very pleased to be at Chicago, which has a very good chairman." Stone laughed it off; he did not take offense.

This lack of communication between mathematicians and physicists was to end soon. By 1977, we all knew about Atiyah and Singer, and then the floodgates came down fast, to the point that an extremely close collaboration between mathematicians and physicists got started and, under the leadership of Ed Witten and others, is ongoing and bearing beautiful fruit to this day.

By the way, on Kap's desk I noticed some work of his on Hopf algebras. I asked him about Hopf algebras, and got the reply, "They are of no relevance whatsoever for physics." I took his word on this, was I ever gullible.

In the wake of our joint work, Kap and I became good friends. This friendship was fueled also by our shared love of music, he was a fine pianist, and I used to sing. For me, the most marvelous part of my collaboration and friendship with Kap was that for the first time I got to see up-close how a great mathematician thinks.

\end{document}